\documentclass[sigconf]{acmart}
\renewcommand\footnotetextcopyrightpermission[1]{}
\AtBeginDocument{%
  }

\usepackage{paralist}
\usepackage{enumitem}
\usepackage{threeparttable}    
\usepackage{multirow}
\usepackage{multicol}
\usepackage{array}
\usepackage{float}
\usepackage{tabularx}
\usepackage{ulem}
\usepackage{listings} %插入代码
\usepackage{graphicx} %插入图片的宏包
\usepackage{float} %设置图片浮动位置的宏包
\usepackage{caption}
\usepackage{pdfpages}
\usepackage{amsmath}
\usepackage{stfloats}
\usepackage{placeins}
\usepackage[linesnumbered,ruled]{algorithm2e}

\begin{document}

\newcommand{\WL}[1]{\textcolor{blue}{[Weile:#1]}}
\newcommand{\LJX}[1]{\textcolor{blue}{[LJX:#1]}}
\newcommand{\ZBN}[1]{\textcolor{blue}{[ZBN:#1]}}
\newcommand{\TODO}[1]{\textcolor{blue}{[TODO:#1]}}

\title{Scaling Neural-Network-Based Molecular Dynamics with Long-Range Electrostatic Interactions to 51 Nanoseconds per Day}

% \author[1]{Author A\thanks{A.A@university.edu}}
% \author[1]{Author B\thanks{B.B@university.edu}}
% \author[1]{Author C\thanks{C.C@university.edu}}
% \author[2]{Author D\thanks{D.D@university.edu}}
% \author[2]{Author E\thanks{E.E@university.edu}}
% \affil[1]{Department of Computer Science, \LaTeX\ University}
% \affil[2]{Department of Mechanical Engineering, \LaTeX\ University}

% \author[*]{test}
% \author[**]{lala}
% \author{Jianxiong Li$^{1,2}$, Tong Zhao$^{1,2}$, Zhuoqiang Guo$^{1,2}$, Shunchen Shi$^{1,2}$, Lijun Liu$^{3}$, Guangming Tan$^{1,2}$, Weile Jia$^{1,2}$*, Guojun Yuan$^{1,2}$, Zhan Wang$^{1,2}$*}
% \affiliation{
%   \institution{1.SKLP, Institute of Computing Technology, Chinese Academy of Sciences}
%   \institution{2.University of Chinese Academy of Sciences, Beijing, China}
%   \institution{3.Osaka University, Osaka, Japan}
%   \city{ } 
%   \country{ }
% }
% \email{lijianxiong@ict.ac.cn, zhaotong@ict.ac.cn, guozhuoqiang20z@ict.ac.cn, shishunchen22z@ict.ac.cn, }
% \email{liu@mech.osaka-u.ac.jp, tgm@ict.ac.cn, jiaweile@ict.ac.cn, yuanguojun@ncic.ac.cn, wangzhan@ncic.ac.cn}

\author{Jianxiong Li}
\affiliation{%
  \institution{SKLP, Institute of Computing Technology, Chinese Academy of Sciences}
  \institution{University of Chinese Academy of Sciences}
  \city{Beijing}
  \country{China}}
\email{lijianxiong20g@ict.ac.cn}

\author{Beining Zhang}
\affiliation{%
  \institution{SKLP, Institute of Computing Technology, Chinese Academy of Sciences}
  \institution{University of Chinese Academy of Sciences}
  \city{Beijing}
  \country{China}}
\email{zhangbeining19@mails.ucas.ac.cn}

\author{Mingzhen Li}
\affiliation{%
  \institution{SKLP, Institute of Computing Technology, Chinese Academy of Sciences}
  \institution{University of Chinese Academy of Sciences}
  \city{Beijing}
  \country{China}}
\email{limingzhen@ict.ac.cn}

\author{Siyu Hu}
\affiliation{%
  \institution{SKLP, Institute of Computing Technology, Chinese Academy of Sciences}
  \institution{University of Chinese Academy of Sciences}
  \city{Beijing}
  \country{China}}
\email{husiyu@ict.ac.cn}

\author{Jinzhe Zeng}
\affiliation{%
  \institution{School of Artificial Intelligence and Data Science, University of Science and Technology of China}
  \institution{Suzhou Big Data \& AI Research and Engineering Center}
  \city{Hefei}
  \country{China}}
\email{jinzhe.zeng@ustc.edu.cn}

\author{Lijun Liu}
\affiliation{%
  \institution{Department of Mechanical Engineering, Graduate School of Engineering, The University of Osaka}
  \city{Suita}
  \country{Japan}}
\email{liu@mech.eng.osaka-u.ac.jp}

\author{Guojun Yuan}
\affiliation{%
  \institution{SKLP, Institute of Computing Technology, Chinese Academy of Sciences}
  \institution{University of Chinese Academy of Sciences}
  \city{Beijing}
  \country{China}}
\email{yuanguojun@ncic.ac.cn}

\author{Zhan Wang}
\affiliation{%
  \institution{SKLP, Institute of Computing Technology, Chinese Academy of Sciences}
  \institution{University of Chinese Academy of Sciences}
  \city{Beijing}
  \country{China}}
\email{wangzhan@ncic.ac.cn}

\author{Guangming Tan}
\affiliation{%
  \institution{SKLP, Institute of Computing Technology, Chinese Academy of Sciences}
  \institution{University of Chinese Academy of Sciences}
  \city{Beijing}
  \country{China}}
\email{tgm@ict.ac.cn}

\author{Weile Jia}
\authornote{Corresponding author}
\affiliation{%
  \institution{SKLP, Institute of Computing Technology, Chinese Academy of Sciences}
  \institution{University of Chinese Academy of Sciences}
  \city{Beijing}
  \country{China}}
\email{jiaweile@ict.ac.cn}


\renewcommand{\shortauthors}{J. Li et al.}

\begin{abstract}

% 包含长程静电力的基于神经网络的分子动力学模拟工具，在短程力基础上，弥补了库伦作用力长尾效应的缺失，可以有效的建模蛋白质折叠，偶极子表面能量等物理现象，在ab inito精度下，扩大了NNMD的应用范围。然而该方法由于涉及复杂神经网络推理以及长程力计算，模拟速度受限。本文我们针对DPLR这一state-of-the-art的工具，提出了全新的基于硬件卸载的FFT计算方法、长程力短程力overlap方法，和基于ring的负载均衡方法。相比baseline工作提升了37倍。在4608个节点上，对216,576个原子的水分子体系，实习拿了37ns/day的模拟速度。
Neural network-based molecular dynamics (NNMD) simulations incorporating long-range electrostatic interactions have significantly extended the applicability to heterogeneous and ionic systems, enabling effective modeling critical physical phenomena such as protein folding and dipolar surface and maintaining \textit{ab initio} accuracy. However, neural network inference and long-range force computation remain the major bottlenecks, severely limiting simulation speed. In this paper, we target DPLR, a state-of-the-art NNMD package that supports long-range electrostatics, and propose a set of comprehensive optimizations to enhance computational efficiency. We introduce (1) a hardware-offloaded FFT method to reduce the communication overhead; (2) an overlapping strategy that hides long-range force computations using a single core per node, and (3) a ring-based load balancing method that enables atom-level task evenly redistribution with minimal communication overhead. Experimental results on the Fugaku supercomputer show that our work achieves a 37x performance improvement, reaching a maximum simulation speed of 51 ns/day.

\end{abstract}



\keywords{
High Performance Computing, Molecular Dynamics, LAMMPS, DeePMD, Computational Science
 }




\maketitle

% \begingroup\renewcommand\thefootnote{*}
% \footnotetext{Corresponding author}

% \endgroup
%\input{section/introduction}
\section{Introduction}

% Weile: start the paper with long range interaction. 

\begin{sloppypar}
    
Accurate modeling of long-range electrostatic interactions is a cornerstone of molecular simulations, particularly in systems where electrostatic forces govern emergent physical and chemical properties. While \textit{ab initio} quantum mechanical methods~\cite{car1985unified}, such as density functional theory (DFT), inherently capture these interactions with high fidelity, their computational cost scales cubically with system size, limiting applications to thousands of atoms and picoseconds simulations~\cite{jia2020pushing}. Neural-network-based molecular dynamics (NNMD), such as SIMPLE-NN~\cite{lee2019simple}, BIM-NN~\cite{bim-nn2017}, HDNNP~\cite{Behler_2014,behler2007generalized,behler2017first}, NequIP~\cite{Batzner2022}, UNEP-v1~\cite{song2024general}, and  SNAP~\cite{THOMPSON2015316}, has emerged as a promising alternative, enabling large-scale simulations with \textit{ab initio} accuracy by learning potential energy surfaces (PES) from quantum mechanical data and achieving linearly scalability. 
%the following sentences are copied from previous text.
For example, Allegro~\cite{musaelian2023learning} employs a local equivariant method, achieving a simulation speed of 114 time-steps/s on a 1 million Ag atom system using 128 A100 GPUs. Currently, the state-of-the-art work on general-purpose supercomputers is DeePMD~\cite{deepmd2024li}, which achieves 1,724 time-steps/s on 0.5 million Cu atom system using 12,000 nodes on Fugaku supercomputer, enabling, for the first time, a 1-microsecond simulation within a week. 
However, these scaling efforts primarily focus on interactions within some fixed cutoff distance, truncating or approximating long-range electrostatic interactions at short distances and neglecting the Coulombic tail of the forces~\cite{Anstine_JPhysChemA_2023_v127_p2417}. 
% This omission leads to systematic errors in predicting properties sensitive to the asymptotic decay of forces, such as interfacial polarization, solvation thermodynamics, and lattice dynamics in ionic crystals.
Although such approximations are sufficient in many cases, in studying complex chemical phenomena, such as heterogeneous bulk phases~\cite{Hart2021}, dispersion interactions~\cite{https://doi.org/10.1002/wcms.30}, hydrogen bonding~\cite{rezac2020non}, long-range charge transfer~\cite{10.1063/1.1590951}, etc, they still suffer accuracy problems due to the truncation of long-range interactions.
\end{sloppypar}

\begin{sloppypar}
Many efforts have been made to integrate the long-range electronic interactions into NNMD to balance computational efficiency and accuracy~\cite{anstine2023machine}. One typical example is Deep Potential Long-Range (DPLR)~\cite{zhang2022deep}. In DPLR, the electrostatic energy is computed from distributions of Gaussian charges centered at ionic sites and valence electron positions, the latter derived from maximally localized Wannier centers~\cite{marzari2012maximally}. The spatial localization of these Wannier centers, which encode the electronic response to the atomic environment, is predicted by a deep neural network trained on \textit{ab initio} data. This approach seamlessly integrates short-range interactions, modeled by the Deep Potential (DP)~\cite{zhang2018deep}, with a scalable long-range electrostatic term, ensuring analytic continuity of forces and the virial tensor and exhibiting high accuracy in modeling different types of water dissociation curves. Besides, PhysNet~\cite{unke2019physnet}, SpookyNet~\cite{Unke2021}, SCFNN~\cite{gao2022self}, etc., have also explored various approaches to compensate for long-range contributions.

\end{sloppypar}

\begin{sloppypar}
Incorporating the long-range electronic interactions in the above physics-induced way to nearsighted short-range NNMD is physical fidelity and thus can enhance the model interpretability. However, the 3D FFT calculation required by methods like Ewald Summation~\cite{ewald1921berechnung}, PME~\cite{darden1993particle} or PPPM~\cite{poisson_ik_hockney1988computer} when solving the electrostatic energy introduces substantial communication overhead in large-scale parallel simulations. This limitation stems from the global data dependencies inherent to Fourier-space computations, which degrade simulation speed and significantly increase time-to-solution.
Additionally, overlapping long-range and short-range force calculations present additional challenges, as long-range force computation can significantly outweigh short-range force computation. Moreover, the uneven distribution of atoms also leads to severe load imbalance across MPI ranks.
    
\end{sloppypar}

To address these challenges and accelerate the simulation speed of NNMD with long-range electrostatic interactions, particularly in scenarios where each core handles only a single atom, this study targets DPLR, a state-of-the-art NNMD package, and introduces a series of systematic, fine-grained optimizations for both computation and communication efficiency to achieve minimal time-to-solution.

\begin{sloppypar}
\begin{itemize}
    \item We propose a communication-efficient 3D-DFT method leveraging hardware offloading on the Fugaku supercomputer. Compared to FFT-MPI~\cite{plimpton2018fftmpi} and heFFTe~\cite{ayala2020heffte} libraries, our method achieves a maximum $2\times$ performance improvement in 3D FFT computations.
    \item We introduce a novel overlap scheme for short-range and long-range force computations which not only enhances the computation resource utilization efficiency but also achieves nearly complete hiding of long-range force calculations.
    \item We further develop a ring-based load balancing algorithm, which achieves atom-level task evenly redistribution while incurring a minimal increase in communication overhead.
    \item Experimental results demonstrate that, compared to the baseline work, our optimizations deliver a maximum $37\times$ performance improvement, achieving a simulation speed of 51 ns/day on 12 computing nodes for a 564-atom water system and maintaining 32 ns/day on 8,400 compute nodes for a 403K-atom water system in large-scale simulation.
\end{itemize}
\end{sloppypar}
Additionally, our optimization strategy not only demonstrates the effectiveness of architecture-specific optimizations but also can be applicable to other NNMD packages and spatial decomposition applications with similar computing patterns.

\begin{figure*}[t]
  \centering
  \includegraphics[scale=0.45]{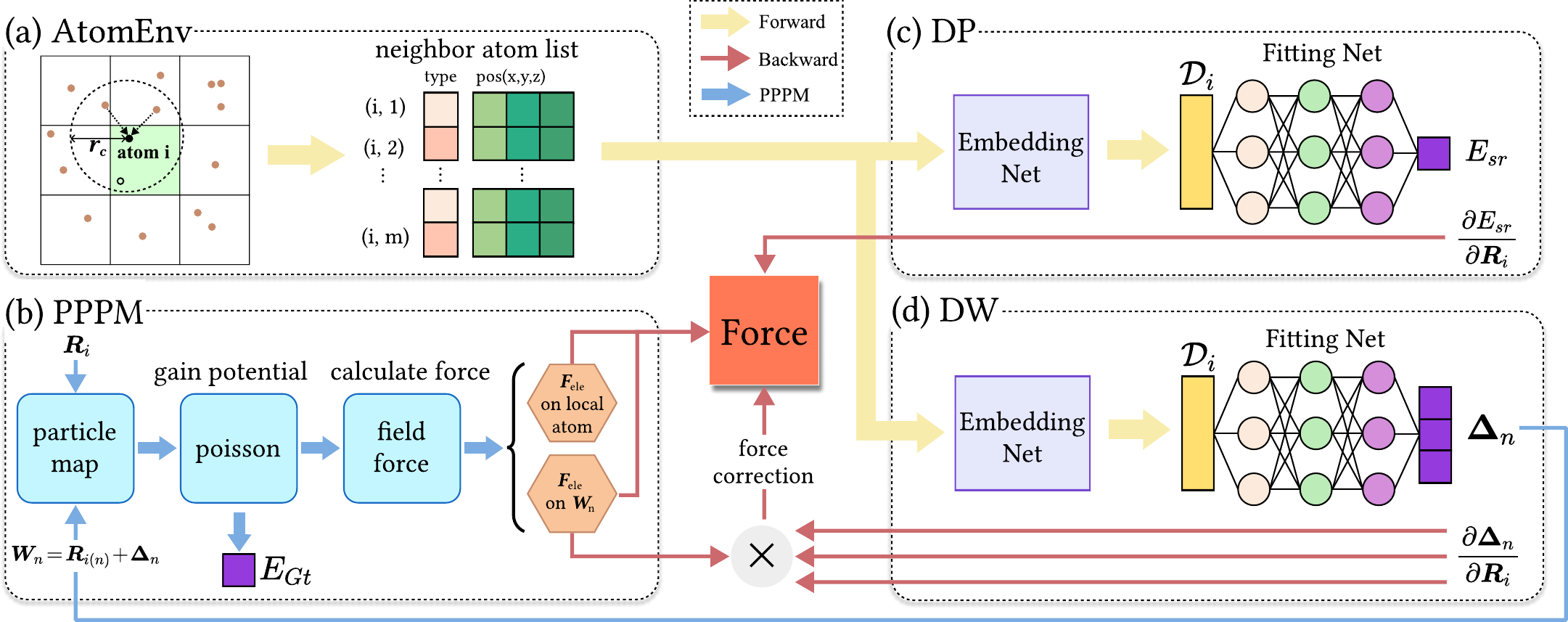}
  \caption{
  A schematic overview of the DPLR. (a) The atom environment of each local atom $i$ is constructed, which is represented as a neighbor list within the cutoff range $r_c$. (b) The electronic energy $E_{Gt}$ and the corresponding electrostatic forces $\boldsymbol F_{ele}$ on local atoms (${\partial E_{Gt}}/{\partial \boldsymbol R_i}$)  and WCs ($\partial E_{Gt} /{\partial \boldsymbol W_{n(i)}}$) are computed using the PPPM method. (c) The short-range energy contribution $E_{sr}$ is obtained through inference using the DP model, with the associated forces computed via backpropagation. (d) The displacements of the WC, denoted as $\boldsymbol \Delta_n$, are predicted by the DW model, and their gradients $-\partial \boldsymbol \Delta_n / \partial \boldsymbol R_i$ are calculated along all three spatial dimensions.
  }
  \label{fig:dplr_computation}
\end{figure*}

\section{Background}

\subsection{Deep Potential Long-Range}

\begin{sloppypar}
Recent advances in neural-network-based molecular dynamics (NNMD) have significantly improved computational efficiency compared to \textit{ab initio} molecular dynamics (AIMD) while retaining \textit{ab initio} accuracy~\cite{Batzner2022, THOMPSON2015316, Unke2021}. However, most existing NNMD packages focus primarily on local interactions within a predefined cutoff range, neglecting the long-tail effect of long-range forces~\cite{Anstine_JPhysChemA_2023_v127_p2417}, which cannot efficiently model phenomena such as electrolyte solutions in contact with electrodes~\cite{scalfi2021molecular}, protein folding problems~\cite{zhou2018electrostatic}, dipolar surfaces~\cite{feller2000molecular}. 

\end{sloppypar}

To address this limitation, the Deep Potential Long-Range (DPLR) model extends the DeepPot-SE model~\cite{zhang2018end} by explicitly incorporating long-range interactions. In the DPLR model, the potential energy surface $E$ is decomposed into two components: 

\begin{equation}
 E = E_{sr} + E_{Gt}
\label{DPLR:E}
\end{equation}

\textbf{$E_{sr}$} denotes the short-range interaction energy derived from the DP model in DPLR, shown in Fig.~\ref{fig:dplr_computation}(c). Notably, the standard DP model implicitly captures both intrinsic short-range contributions and the short-range portion of electrostatic interactions during training. To avoid double counting the electrostatic interactions, DPLR subtracts the electrostatic energy from the total energy before training the DP model. 

Within the DP, an \textit{embedding net} extracts the atomic descriptors $\mathcal{D}_i$ from the neighbor list of atom $i$. The descriptors encode the physical symmetries of the molecular systems and are subsequently passed through a \textit{fitting network}, which is implemented as a three-layer fully connected neural network, to predict the short-range energy $E_{sr}$. The corresponding short-range forces, $\boldsymbol F_{sr}=\frac{\partial E_{sr}}{\partial \boldsymbol R_i}$, are computed via backpropagation, where $\boldsymbol R_i$ represents the atomic coordinates.
% preserve the underlying physical symmetries.
 % environment matrix $\tilde R_i$
 % perserve the translational, rotational, and permutation invariances inherent of the molecular systems~\cite{}.

\textbf{$E_{Gt}$} stands for the long-range electrostatic energy. DPLR represents long-range electrostatic interactions via spherical Gaussian charge distributions placed at well-defined ionic and electronic sites. The latter, referred to as Wannier centroids (WCs), are defined by the averages of the positions of the maximally localized Wannier centers~\cite{marzari2012maximally} binding to specific atoms~\cite{zhang2020deep}. The long-range electrostatic energy is efficiently computed using the particle-particle-particle-mesh (PPPM) method~\cite{poisson_ik_hockney1988computer} in Fourier space, as illustrated in Fig.~\ref{fig:dplr_computation}(b).

\begin{equation}
E_{Gt} = \frac{1}{2 \pi V} \sum_{m \neq 0, m \leq L} \frac{\exp(-\pi^2 m^2 / \beta^2)}{m^2} S^2(m) ,
\label{DPLR:EGt}
\end{equation}

\begin{equation}
S(m) = \sum_{i} q_i e^{-2\pi \mathrm{i} m \boldsymbol R_i} + \sum_{n} q_n e^{-2\pi \mathrm{i} m \boldsymbol{W}_n},
\label{DPLR:Sm}
\end{equation}
where
the $\beta$ determines the Gaussian width, L is the cutoff radius in Fourier space, $\mathrm{i} = \sqrt{-1}$, $q_i$ and $q_n$ denote ionic and electronic charges, $\boldsymbol R_i$ represents ionic coordinates, and $\boldsymbol W_n$ corresponds to the $n$-th WC. Owing to the tight binding between WCs and their nearest atoms, $\boldsymbol W_n$ can be determined by

\begin{equation}
\boldsymbol W_n = \boldsymbol R_{i(n)} + \boldsymbol \Delta_n,
\label{DPLR:Wn}
\end{equation}
where $\boldsymbol \Delta_n$ denotes the displacement of $\boldsymbol W_n$ from $\boldsymbol R_i$, which can be well modeled via the Deep Wannier (DW) model.
The DW model, which is illustrated in Fig.~\ref{fig:dplr_computation}(d), has a similar structure to the DP model, which leverages the binding atoms along with their neighbor list to predict the $\boldsymbol \Delta_n$ via model inference.

Then, the force on atom $i$ can be derived by

\begin{equation}
\boldsymbol F_i = -\frac{\partial E}{\partial \boldsymbol R_i} = -\frac{\partial E_{sr}}{\partial \boldsymbol R_i} - \frac{\partial E_{Gt}}{\partial \boldsymbol R_i} - \sum_n \frac{\partial E_{Gt}}{\partial \boldsymbol W_n} \frac{\partial \boldsymbol W_n}{\partial \boldsymbol R_i},
\label{DPLR:F_inter}
\end{equation}
Incorporating the WC-atom binding relation shown in Equ.~\ref{DPLR:Wn}, the force expands to
\begin{equation}
\boldsymbol F_i = -\frac{\partial E_{sr}}{\partial \boldsymbol R_i} - \frac{\partial E_{Gt}}{\partial \boldsymbol R_i} - \frac{\partial E_{Gt}}{\partial \boldsymbol W_{n(i)}} - \sum_n \frac{\partial E_{Gt}}{\partial \boldsymbol W_n} \frac{\partial \boldsymbol \Delta_n}{\partial \boldsymbol R_i}
\label{DPLR:F_final}
\end{equation}

While the inclusion of long-range interactions enables DPLR to achieve higher accuracy than the standard DP model, it also introduces additional computational complexity. Specifically, the use of the PPPM method, which relies on 3D-FFT computations, results in substantial communication overhead in large-scale simulations. Furthermore, each time-step requires two neural network inferences, which further increases computational cost. Consequently, enhancing the efficiency of DPLR in large-scale simulations remains a key challenge.

\subsection{Fugaku supercomputer}

\begin{sloppypar}
    
The Fugaku supercomputer, located in the RIKEN center in Japan, represents a landmark achievement in ARM-based high-performance computing. It comprises 158,976 computing nodes with a peak performance of 537 PFLOPS, currently ranking 6th on the TOP500 list~\cite{top500} and 4th on the HPL-MxP benchmark~\cite{hpl-mxp}. 

\begin{figure}[tb]
  \centering
  \includegraphics[scale=0.4]{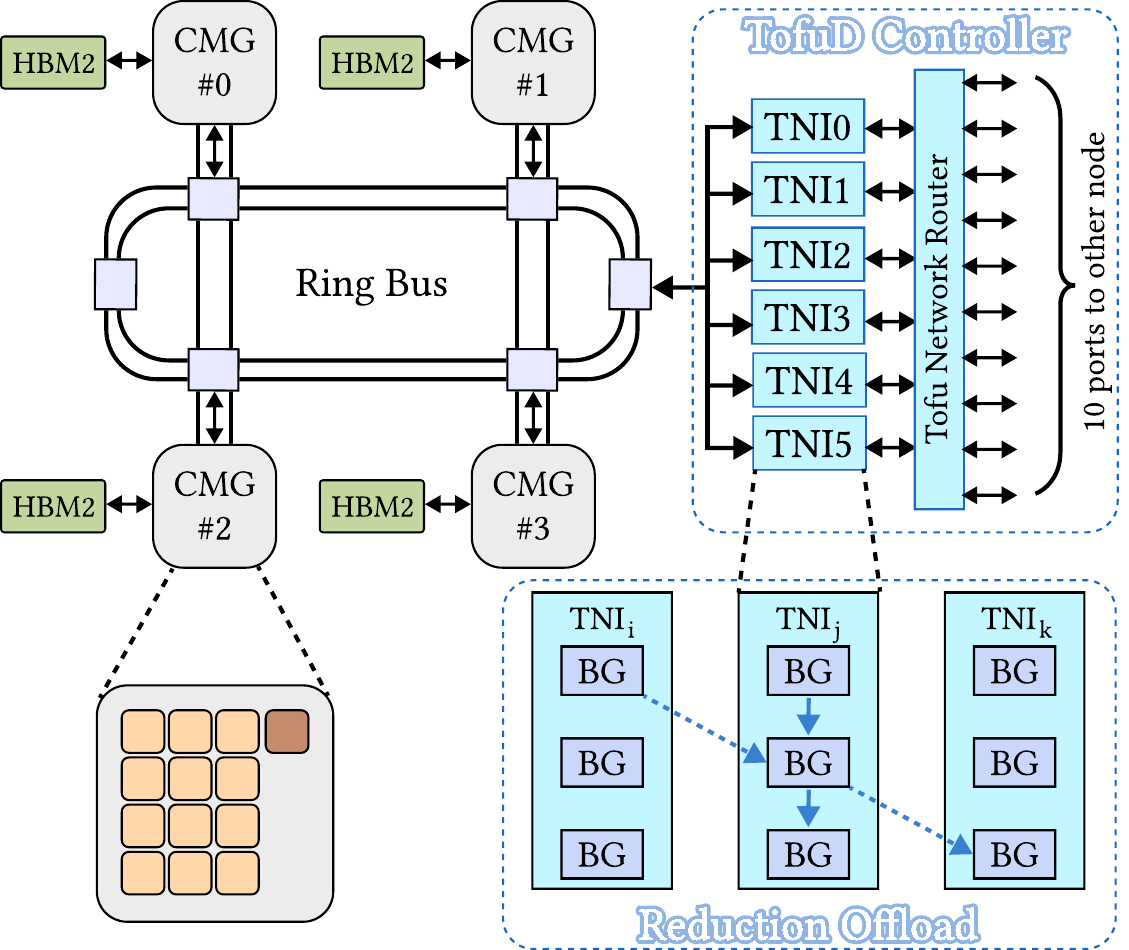}
  \caption{
  The architecture of the A64FX processor. The processor is organized into four Core Memory Groups (CMGs), coupled with a TofuD controller responsible for inter-node communication. The TofuD controller integrates six TNIs and ten network ports, enabling direct connections to ten neighboring nodes in the 6D torus/mesh topology. Fugaku supports hardware offloading for reduction operations through the BGs embedded within the TNIs, which can be flexibly configured into reduction chains and enable low-latency data aggregation.
  }
  \label{fig:fugaku_cpu}
\end{figure}

Each node of Fugaku integrates an A64FX processor featuring 52 cores organized into four NUMA domains (also called Core Memory Groups (CMGs))~\cite{fugaku_cpu}, as shown in Fig.~\ref{fig:fugaku_cpu}. Within each CMG, one core is dedicated to the operating system and I/O while the remaining 12 cores handle computational tasks. Each CMG is equipped with an 8 GB HBM2 memory providing 256GB/s bandwidth. The processor implements 512-bit Scalable Vector Extensions (SVE) with dual SIMD pipelines, enabling each core to execute up to 32 double-precision floating-point operations per cycle.
\end{sloppypar}

The inter-node communication of Fugaku is managed by the Tofu Interconnect D (TofuD)~\cite{tofuD_network}, which employs a 6D torus/mesh topology with 10 ports per node. The high-dimensional network topology enables efficient mapping of MPI ranks to underlying node layout, significantly enhancing the communication efficiency of domain-decomposition applications such as MD simulations. 

Each node is equipped with six Tofu Network Interfaces (TNIs), enabling simultaneous handling of six messages. 
Fugaku further supports network function offloading for reduction operation with extremely low latency. Each TNI contains 48 hardware units, named Barrier Gates (BGs), including 16 Start/End BGs responsible for initiating and terminating communication, and 32 Relay BGs dedicated to relaying and aggregating data. These BGs can be flexibly configured to form reduction chains, enabling low-latency operations such as summation, bitwise OR, and bitwise AND over six 64-bit unsigned integers or three double-precision values. For example, an allreduce operation across 10,000 nodes can be completed in as little as 7 microseconds using a binary tree configuration.

These advanced architectural features make Fugaku exceptionally well-suited for optimizing communication-intensive scientific applications and offer valuable opportunities to enhance the performance and scalability of DPLR in large-scale MD simulations.

\section{Optimization}

This section introduces our main optimizations in detail, including the hardware-offloaded 3D FFT computation, the overlapping strategy for long-range and short-range force calculations, and the ring-based load balancing method.

\subsection{Hardware-offloaded 3D FFT}

DPLR employs the PPPM method with the Poisson-IK algorithm~\cite{poisson_ik_hockney1988computer} to calculate electrostatic forces, where the main computational bottleneck arises from the 3D FFT. Specifically, one forward and three inverse 3D FFTs are required.

Two approaches are commonly adopted for 3D-FFT calculations in MD simulations. The first performs 3D-FFT directly across all MPI ranks arranged in a 3D decomposition. The method involves \textit{brick-to-pencil} or \textit{brick-to-slab} redistribution to facilitate 1D- or 2D-FFT, followed by \textit{pencil-to-pencil} or \textit{slab-to-pencil} transpose to complete the FFT across all dimensions~\cite{gholami2015accfft, ayala2020heffte, plimpton2018fftmpi}. The second approach partitions MPI ranks into two separate groups, one dedicated solely to long-range interaction calculations~\cite{hess2008gromacs}. Although the latter generally offers superior performance on large scales, communication overhead still remains a significant bottleneck.

Thus, we propose a novel communication-efficient approach that utilizes hardware offloading for 3D-FFT, termed utofu-FFT. Recall the standard Discrete Fourier Transform (DFT), defined in ~\ref{equ:DFT}, where $\omega_{N}^{kn}=e^{-\mathrm i\frac{2\pi}{N}nk}$, $e^{\mathrm ix}=cosx+\mathrm i\cdot sinx, \ \mathrm i=\sqrt{-1}$. 

\begin{equation} \label{equ:DFT}
    X(k)=\sum_{n=0}^{N-1}\omega_{N}^{kn} x(n), k=0,1,...,N-1,
\end{equation}
% The values in $K$-space, represented as $X(k)$, are obtained via a summation of products between real-space grid values $x(k)$ and twiddle factors along each dimension. This operation can be reformulated as a matrix multiplication $X=F_N x$, that is:  
which can be represented as matrix-vector multiplication $X = F_N x$:

\begin{equation} \label{equ:DFT_Matrix}
\begin{bmatrix}
X(0) \\
X(1) \\
\vdots \\
X(K-1)
\end{bmatrix}
=
\begin{bmatrix}
\omega_N^{0 \cdot 0} &  \cdots & \omega_N^{0 \cdot (N-1)} \\
\omega_N^{1 \cdot 0} &  \cdots & \omega_N^{1 \cdot (N-1)} \\
\vdots & \ddots & \vdots \\
\omega_N^{(K-1) \cdot 0} &  \cdots & \omega_N^{(K-1) \cdot (N-1)}
\end{bmatrix}
\begin{bmatrix}
x(0) \\
x(1) \\
\vdots \\
x(N-1)
\end{bmatrix}
\end{equation}

\begin{figure}[htb]
  \centering
  \includegraphics[scale=0.2]{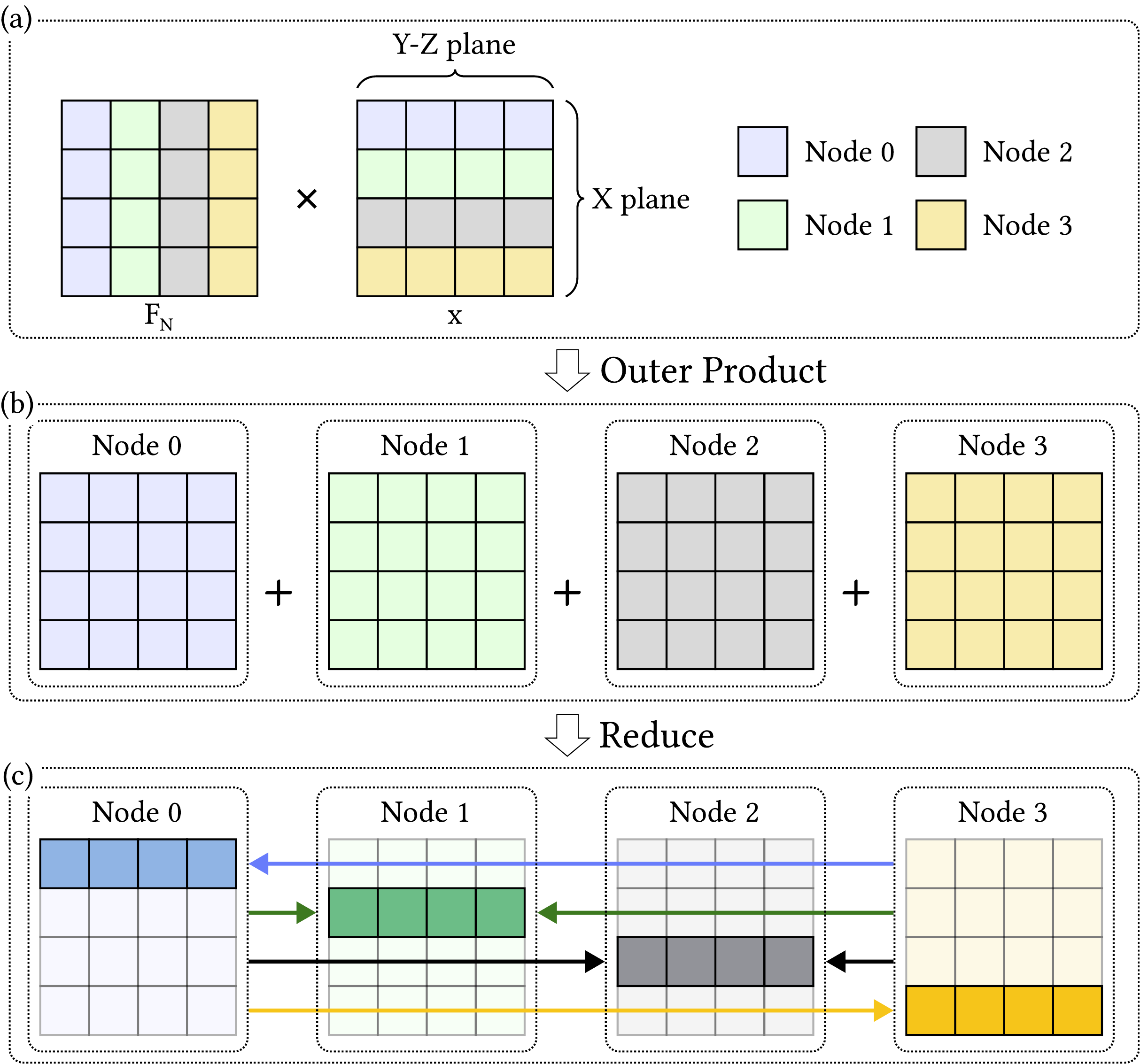}
  \caption{
  The utofu-FFT computation process in X-dimension (a) Each MPI rank contains only a subset of real-space grid. 
  (b) Each MPI rank independently performs partial DFT computations on its assigned data segment. (c) A reduction operation is employed across MPI ranks to aggregate partial results and reconstruct the corresponding subset of the K-space data.
  }
  \label{fig:fft_compute}
\end{figure}

\begin{sloppypar}
Here, $F_N$ is the twiddle factor matrix, $x$ is the input in real space, and $X$ is the transformed data in $K$-space. Since each MPI rank holds only a subset of $x$, local computation becomes $\tilde X = F_{N[:,J]}x_{[J,:]}$, where $J$ denotes the index set of selected columns or rows. To recover the final $K$-space data, MPI ranks of the given dimension must sum up their partial results, as shown in Fig.~\ref{fig:fft_compute}, where the reduction operation can be offloaded to Fugaku's hardware.    
\end{sloppypar}

\begin{figure}[htb]
  \centering
  \includegraphics[scale=0.5]{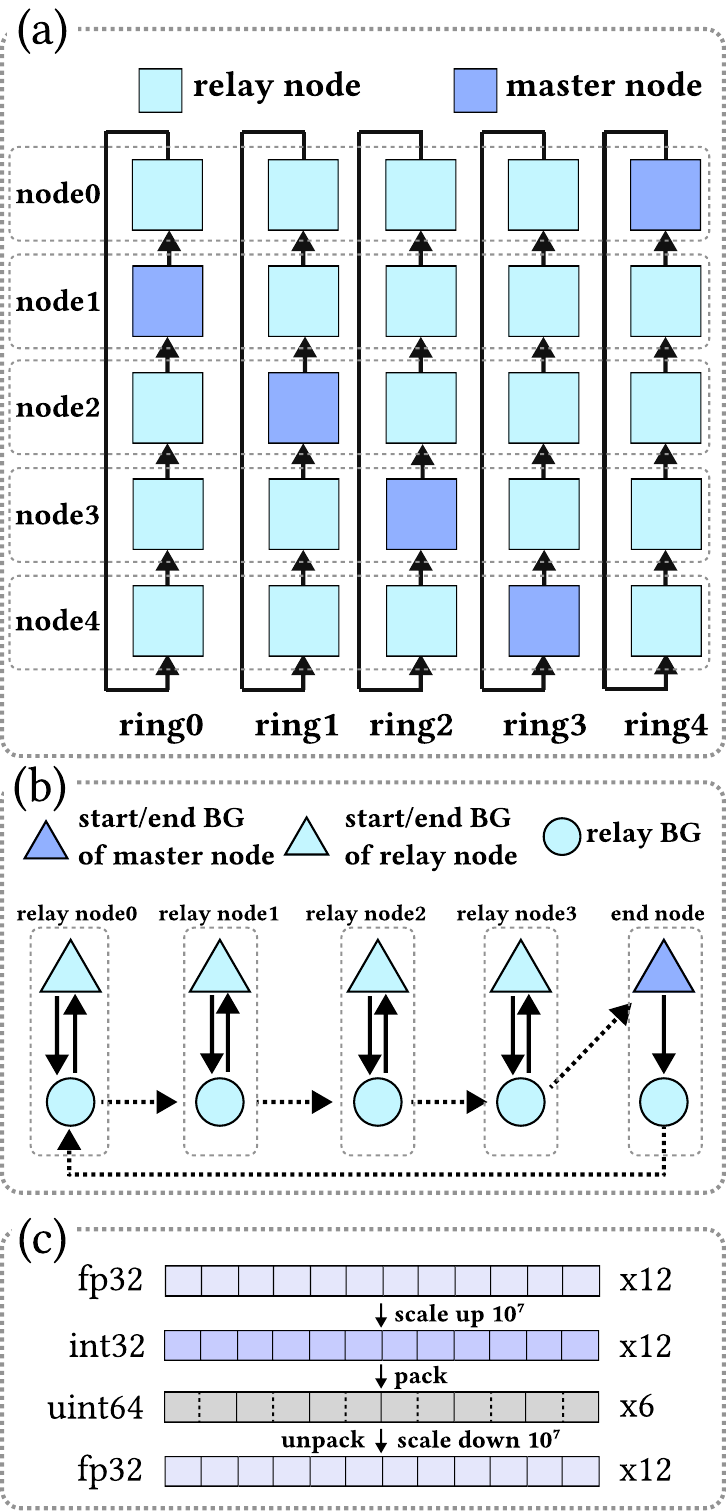}
  \caption{
  Hardware-offloaded reduction process for FFT. (a) For a given dimension, nodes are organized into \textit{n}-rings, each node serving as the master in one ring, responsible for initiating communication and receiving the final reduction result. (b) The reduction communication starts from the node's Start/End BGs. The data is aggregated along the pre-configured reduction chain and then returned to the master node. (c) Data Quantization Method. The original floating-point data is scaled up by a factor of $10^7$ and converted to \textit{int32}. Each two of them is then packed into a single \textit{uint64} for reduction communication. The reduction results are decoded and scaled down to retrieve the final result.
  }
  \label{fig:ring_fft}
\end{figure}

To achieve efficient offloading for distributed 3D FFT computations, well-designated reduction chains of BG configuration are essential. Each BG features two input ports and two output ports. One input port receives data packets from an external TNI, while the other accepts signals from other local BG. Upon receiving both inputs, the BG performs a reduction operation and forwards the result as a data packet to an external BG while simultaneously sending a signal to another local BG. Notably, each BG can be configured to use only one input and one output port at a time, leaving the other idle.

Given the need for reduction along all three dimensions, the six TNIs are partitioned into three groups, each comprising two TNIs. Each TNI allocates 24 BGs: 12 for Start/End and 12 for Relay. Due to constraints of the underlying protocol stack, some BGs must be reserved for other barrier operations, limiting the availability for FFT reduction.

Our reduction chain strategy arranges nodes into ring topologies per dimension. For $n$ nodes in a dimension, we form $n$ rings, with each node serving as the master for one ring, shown in Fig.~\ref{fig:ring_fft}(a). The master node is responsible to initiate the reduction operation and get the final result. Taking Ring 4 as an example, where Node 4 serves as the master node, the reduction chain is configured as Fig.~\ref{fig:ring_fft}(b). The corresponding reduction communication proceeds as follows:

\begin{enumerate}
    \item An MPI rank in each node initiates reduction via a communication primitive. The Start/End BGs receive the command and send a signal containing the pending value to their corresponding Relay BGs.
    \item Node 4’s Relay BG sends a data packet containing the pending value to Node 0’s Relay BG.
    \item Upon receiving both inputs, Node 0's Relay BG performs local reduction and forwards the result to Node 1's Relay BG while signaling completion locally.
    \item The chain continues Node 0 → 1 → 2 → 3.
    \item Node 3’s Relay BG sends the final result back to Node 4’s Start/End BG, where the master MPI rank of Node 4 retrieves it.
\end{enumerate}

The ring-topology reduction chain enables highly efficient summation of distributed FFT data. Nonetheless, the hardware offloading approach faces a key constraint: each reduction operation supports only three \textit{double} or six \textit{uint64} values, and the reduction operation within the same chain must be fully completed before initiating the next one. To achieve the fastest time-to-solution, even with a minimal subdomain per node, a minimum of $4 \times 4 \times 4 = 64$ grid points is required to ensure numerical accuracy, leading to $2 \times 64$ values per dimension (accounting for real and imaginary parts). Even with \textit{uint64} quantization, 22 times reduction operations are required, introducing significant overhead.

To alleviate this, we propose an optimized data quantization scheme that utilizes \textit{int32} quantization, shown in Fig.~\ref{fig:ring_fft}(c). Since most values lie within the range $[-1,1]$, we scale them by $1 \times 10^7$ to preserve precision, then convert them to \textit{int32}. Then, two \textit{int32} values are packed into a single \textit{uint64}, allowing each reduction operation to process 12 values. This method reduces the number of reductions per dimension from 22 to 11. Given that each reduction takes only a few microseconds, a full 3D-FFT can be completed within hundreds of microseconds.

Since up to 24 independent chains can be instantiated in each dimension (12 chains for each TNI limited by BG numbers), multiple reduction chains per node can be employed to further accelerate computation if the node number in a dimension is fewer than 12.

\subsection{Long-range and Short-range Force Overlap}

\begin{figure}[t]
  \centering
  \includegraphics[scale=0.24]{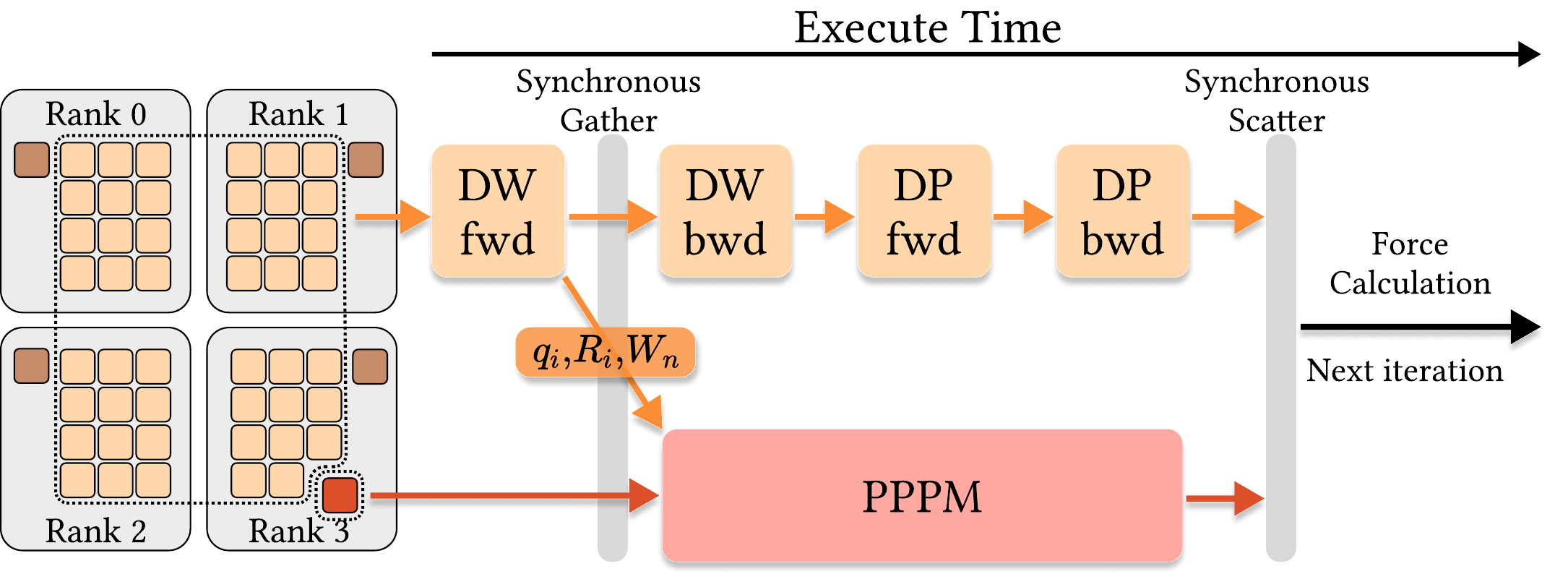}
  \caption{
    The long-range and short-range force overlap strategy. One core in Rank 3 is dedicated to PPPM computations, while the others handle the DW and DP model calculations. 
  }
  \label{fig:sr_lr_overlap}
\end{figure}

A commonly adopted strategy for overlapping long-range and short-range force computations is to partition MPI ranks into two groups, dedicating one group exclusively to long-range interactions. This method significantly reduces the number of MPI ranks involved in FFT communication, thereby improving efficiency. However, it suffers from suboptimal resource utilization, as a substantial portion of nodes, typically around one-quarter, must still be reserved for PPPM tasks. Additionally, in scenarios leveraging the utofu-FFT, the limited participation of nodes in FFT operations leads to the under-utilization of hardware offloading resources.

Thus, we propose an enhanced overlap scheme in which each node employs four MPI ranks, dedicating one core in Rank 3 exclusively for PPPM computation while utilizing the remaining 47 cores for the DP and DW model inference, as shown in Fig.~\ref{fig:sr_lr_overlap}. Since PPPM needs the WC positions, the remaining 47 cores first calculate the forward phase of the DW model to get the $\boldsymbol \Delta_n$. After each MPI rank fixes the WC positions, Rank 3 performs a gather operation to collect particle and WC positions and charges from other MPI ranks within the same node. After finishing the PPPM computation, Rank 3 redistributes the electrostatic forces back to the respective MPI ranks via a scatter operation.

This intra-node overlapping scheme reduces the number of cores involved in communication-intensive long-range force computation to just 1/48 per node. Meanwhile, it reduces the number of MPI ranks involved in the FFT calculation by three quarters while maximizing the utilization of BG hardware resources.

\subsection{Ring-Based Load Balance}

\begin{figure}[t]
  \centering
  \includegraphics[scale=0.40]{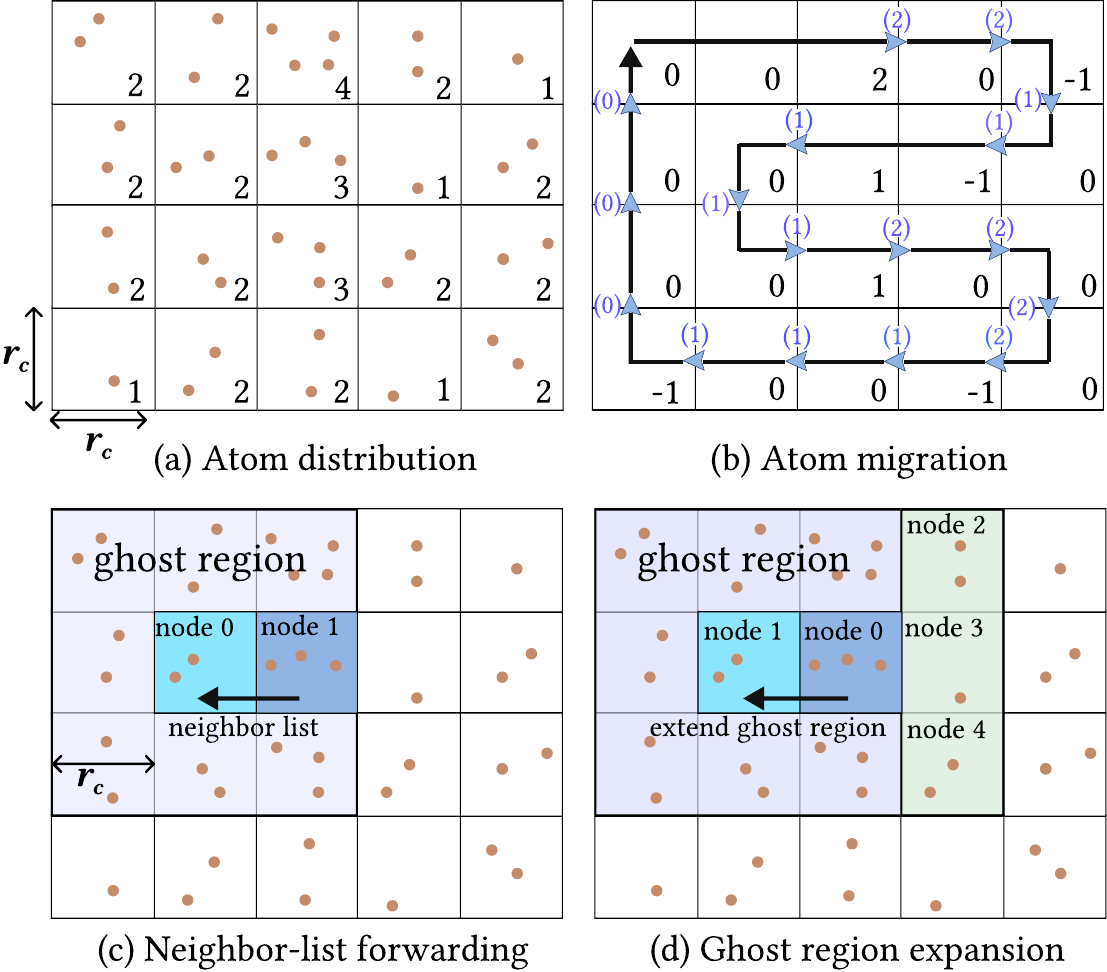}
  \caption{
  Ring-Based Load Balancing Strategy. (a) Initial Atomic Distribution: In the illustrated scenario, under an ideal load-balanced condition, each MPI rank should handle two atoms. (b) Ring Topology Connection: MPI ranks are organized in a ring topology. Each MPI rank determines the number of atoms to be migrated. The black numbers represent the deviation of the atom count in each MPI rank from the ideal load-balanced state, while the blue numbers indicate the number of atoms to be migrated. (c) Neighbor-list forwarding method. The donor MPI rank packages the atoms to be migrated along with their neighbor lists and sends them to the downstream MPI rank. The downstream rank processes the received atoms and returns the results upon completion. (d) Each MPI rank extends the ghost region in the opposite direction of its upstream MPI rank, allowing for the computation of atoms from the upstream MPI rank.
  }
  \label{fig:ring_load_balance}
\end{figure}

The ideal objective of load balancing is to ensure that each MPI rank, or even each core, handles an equal number of atoms. Currently, there are three primary load balancing strategies, each of which faces certain limitations.

% such as increased communication overhead or imperfect load balancing.

\begin{itemize}
    \item Non-uniform spatial decomposition, which is widely adopted in domain-decomposed applications. However, it often introduces additional communication overhead and does not fully achieve atom-level load balancing.
    \item Intra-node atomic balancing~\cite{deepmd2024li}, which performs effectively in systems with uniform density. Nonetheless, particle motion over time can still lead to an atomic imbalance between computing nodes.
    \item Atom swap method~\cite{cerebras2024breaking}, which have demonstrated efficiency on the Cerebras system. Nevertheless, the long-distance communication hops will introduce substantial overhead on pseudo-3D torus topology supercomputers.
\end{itemize}

% To achieve the aforementioned goal while minimizing additional overhead, we propose a novel ring-based load balance method. 
The load imbalance issue in MD simulations is due to the suboptimal spatial decomposition strategies.
The geometry-based partition does not account for the actual computational workload in each subdomain, causing some MPI ranks assigned significantly more atoms than others and leading to uneven computational load. To address this issue, inspired by the atom swap method, we propose a shift from spatial partition to atomic-centric partition, where computational tasks are redistributed at the granularity of individual atoms. Unlike the atom swap method, we adopt a global migration manner.

In our approach, we still take the spatial decomposition at the MPI rank level to ensure spatial locality at the beginning. Assume the total number of atoms is $N_{all}$ and there are $nranks$ MPI ranks. The target number of atoms per MPI rank after load balancing is $N_{goal}$, where $N_{goal} = N_{all} / nranks$, which is ideally set to an integer multiple of the number of cores per MPI rank. If the local atom number $N_{local}$ of an MPI rank is greater than $N_{goal}$, the computation tasks of the excess atoms have to be migrated to other MPI ranks. Since these migrated atoms do not initially belong to the new subdomain, the recipient MPI rank must adjust its ghost region and communication lists based on the newly received atom positions. The newly added ghost region is the primary extra overhead. To minimize this, the ghost region of the migrated atom should overlap with the recipient MPI rank’s ghost region as much as possible. The most straightforward way is to ensure atoms only migrate to immediate neighboring MPI ranks.

To this end, we propose a ring-based atom migration algorithm, shown in Fig.~\ref{fig:ring_load_balance}. The key idea is to form all MPI ranks into a directed ring, where each MPI rank only exchanges atoms with its immediate neighbors in the ring, shown as Fig.~\ref{fig:ring_load_balance}(b). Each MPI rank receives excess atoms from its upstream neighbor and sends its excess atoms to its downstream neighbor. 

% After each Exchange stage of Lammps, which performs dozens of time-steps once to exchange the fly-away atoms, 
The MPI ranks launch a global allgather operation to collect the atom numbers, denoted as $N_{m}[i]$ for each MPI rank $i$. Then, each MPI rank determines the number of atoms to be sent, denoted as $N_{s}[i]$, which can be derived via Algorithm~\ref{alg:load_balance}.

\begin{algorithm}[tb]
    \caption{Ring-based Load Balancing Migration Algorithm}
    \label{alg:load_balance}
    \KwData{
        upstream,
        downstream,
        $N_{local}$,
        $N_{goal}$,
        nranks
    }
    \KwResult{
        $N_s$
    }
    \BlankLine

    % \tcp{Compute per-rank load deviation}
    % \For{$i \leftarrow 0$ \KwTo $nranks-1$}{
    %     $N_s[i] \leftarrow N_{local}[i] - N_{goal}$ \;
    % }
    $cur \leftarrow 0$\;
    set $N_s$ to 0\;

    % \tcp{Two-pass migration via ring propagation}
    \For{$iter \leftarrow 1$ \KwTo $2$}{
        \For{$i \leftarrow 0$ \KwTo $nranks-1$}{
            $pre \leftarrow upstream[cur]$\;
            $nxt \leftarrow downstream[cur]$\;
            \tcp{Update the send atom number}
            $N_s[cur] \leftarrow N_{goal}[cur] - N_{local}[cur] + N_s[pre]$\;

            \tcp{Send number should not exceed local atom number and not lower than zero}
            \If{$N_s[cur] < 0$}{
                $N_s[cur] \leftarrow 0$\;
            }
            \If{$N_s[cur] > N_{local}[cur]$}{
                $N_s[cur] \leftarrow N_{local}[cur]$\;
            }
            $cur \leftarrow nxt$\;
        }
    }
    
\end{algorithm}

Fig.~\ref{fig:ring_load_balance} illustrates an example where $N_{goal}$ equals $2$. The initial atom distribution is shown in Fig.~\ref{fig:ring_load_balance}(a), while Fig.~\ref{fig:ring_load_balance}(b) shows the number of atoms to be migrated to the downstream MPI rank. With the ring-based load balance algorithm, the global load balancing can be achieved by migrating a minimal number of atoms, with each atom moving only a single hop to its immediate neighboring MPI rank. The allgather communication operation only needs to be performed once every several dozen time-steps, resulting in negligible overhead. 

Building upon this algorithm, we further propose two strategies for migrating computational tasks.

\textbf{Neighbor-list forwarding}. As illustrated in Fig.~\ref{fig:ring_load_balance}(c), the straightforward way is to make the upstream MPI rank to pack the migrating atoms along with their neighbor lists and then send the data to the downstream MPI rank. The recipient MPI rank subsequently performs the force calculations locally and returns the result to its upstream MPI rank. This method introduces additional twice synchronized message passing. Moreover, the volume of data sent depends on both the number of migrated atoms and the size of their neighbor lists.

\textbf{Ghost region expansion}. As shown in Fig.~\ref{fig:ring_load_balance}(d), since the migrated atoms originally reside in the ghost region of the downstream MPI rank, eliminating the need for another explicit transmission, the downstream MPI rank only needs to incorporate the ghost region of the upstream node into its own ghost region for calculation. Given that most of the ghost region is overlapped, the expansion of the ghost region remains minimal. This method preserves the conventional force computation workflow while introducing only a slight communication overhead.

% The communication overhead can be accurately estimated using the LogP model~\cite{logpculler}, allowing user to assess whether the performance gains from load balancing outweigh the additional communication costs.

The ring-based load balance method effectively mitigates load imbalance across nodes. Meanwhile, since the ghost region expansion method can leverage the original atom distribution and incur no synchronization overhead, it achieves higher efficiency in scenarios with minimal per-node atom counts.

The ring topology over the 3D-distributed MPI ranks is constructed by the serpentine scanning algorithm. To reduce communication hops, we leverage the \textit{mpi-ext} library to obtain the node coordinate and achieve an optimized mapping between MPI ranks and the physical node topology, which further improves communication efficiency.

\subsection{Other Optimizations}

\subsubsection{Node-level task division}

DPLR utilizes the LAMMPS framework for MD simulation. The original LAMMPS code performs spatial decomposition, local atomic computations, and ghost atom communications at the MPI rank level. However, when each MPI rank is assigned a very small spatial domain, each MPI rank may need to communicate with two layers of neighboring MPI ranks to acquire ghost atoms. When using the ring-based load balance method, each MPI rank must interact with even more additional MPI ranks, leading to significant communication overhead. Additionally, some MPI ranks may contain very few atoms, fewer than the number required to be migrated by the ring-based method, rendering the ring-based load balancing approach inapplicable.

To overcome these limitations, we leverage insights from prior research~\cite{deepmd2024li} and propose a refined task decomposition methodology that shifts the workload partitioning granularity from individual MPI ranks to per-node allocation. Specifically, we employ an allgather operation within each node, enabling every MPI rank to obtain complete information of all local atoms within the node. Based on this, the communication pattern shifts from being MPI-rank-centric to node-centric, where each node is responsible for exchanging ghost atoms with neighboring nodes. To efficiently handle this, we distribute the communication tasks across all MPI ranks within a node. In particular, each MPI rank first collects ghost atom data from its assigned neighboring nodes and then broadcasts this information to all ranks within the same node, ensuring that every MPI rank has access to the complete set of local and ghost atoms of the node.

With this transformation, all threads (cores) within a node can evenly distribute computational tasks. Additionally, in the ring-based load balance method, atomic migration is performed at the node level rather than the rank level, preventing cases where individual MPI ranks lack sufficient atoms for migration.

\subsubsection{Model Inference Optimization}

When performing MD simulations, DPLR utilizes a neural network framework for model inference to compute atomic forces and energies. However, our tests revealed that the framework exhibits low efficiency when the workload is minimal. Specifically, many automatically generated kernels in the gradient computation process are redundant, leading to excessive unnecessary computations. Additionally, the framework incurs substantial initialization and management overhead, the actual execution time of all computational kernels accounts for less than half of the total inference time. Furthermore, certain built-in operators, such as \textit{tanh} and \textit{matmul}, demonstrate suboptimal efficiency.  

To address these challenges, we restructure the DPLR code based on the computational workflow and implemented a framework-free code. Additionally, we applied fine-grained kernel optimizations and kernel fusion, significantly enhancing computational efficiency.

\section{Evaluation}

We use the water system as a representative example for our experiments. Water is a widely adopted benchmark system for training and evaluation in molecular dynamics~\cite{jia2020pushing, guo2022extending, liang2025polarizable}. Due to the intricate balance between weak non-covalent intermolecular interactions, thermal (entropic) effects, and nuclear quantum effects~\cite{ko2019_water, distasio2014_water, chen2017_water}, the water system poses long-standing challenges in accurate simulations. In DPLR, the WC of a water molecule is binding to the oxygen atom. 

The cutoff range for each atom is set to 6 \AA. The number of neighbor atoms for water and hydrogen atoms is 46 and 92, respectively. The \textit{fitting net} structure for both the DP model and DW model is (240, 240, 240). The time-step scale is set to 1 fs. The neighbor list is built with a skin distance of 2 \AA~ and is updated every 50 time steps. Simulations are performed under the NVT ensemble, with the initial temperature set to 300 K.

\begin{table}[b]
  \begin{center}
    \caption{Error of the energy and force for one time-step}
    \label{tab:error_energy}
    \begin{threeparttable}
    \begin{tabular}{c c c c c c}\hline
      \textbf{Precision}  & \begin{tabular}[c]{@{}c@{}}\textbf{Error in energy} \\ {[}eV/atom{]}\end{tabular}  &  \begin{tabular}[c]{@{}c@{}}\textbf{Error in force} \\ {[}eV/A{]}\end{tabular} \\  \hline
      Double(32x32x32)                & $3.7\times10^{-4} $              & $5.3\times10^{-2} $  \\ \hline
      Mixed-fp32(32x32x32)                 & $3.7\times10^{-4} $      & $5.3\times10^{-2} $  \\  \hline
      Mixed-int0(12x18x12)            & $3.5\times10^{-4} $             & $5.3\times10^{-2} $ \\ \hline
      Mixed-int1(10x15x10)            & $3.9\times10^{-4} $             & $5.3\times10^{-2} $ \\ \hline
      Mixed-int2(8x12x8)            & $3.7\times10^{-4} $             & $5.3\times10^{-2} $ \\ \hline
    \end{tabular}
    \begin{tablenotes}    
        \footnotesize            
        \item[1] The error is compared with AIMD result.   
        \item[2] The number in parentheses is the 3D-FFT grid.
    \end{tablenotes}         
  \end{threeparttable}
  \end{center}
\end{table}

\begin{figure}[b]
  \centering
  \includegraphics[scale=0.4]{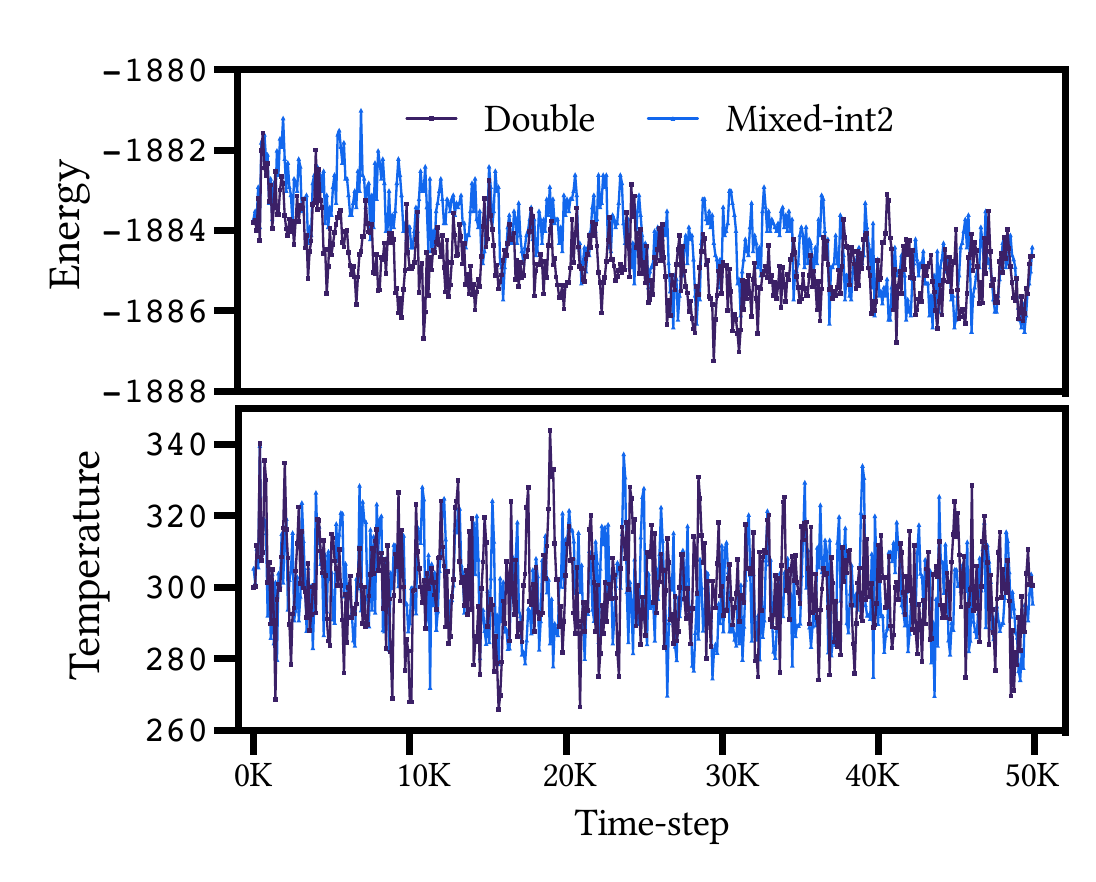}
  \caption{
  The energy and temperature for 50K time-steps under double and mixed-int2 precision.
  }
  \label{fig:press}
\end{figure}

All experiments are conducted on the Fugaku supercomputer using the tsdc-1.2.41 runtime environment, which includes compiler tool-chains, communication primitive and BLAS libraries. The TensorFlow version is 2.2. For FFT computations, we adopt FFTW version 3.3.10 as the backend. Two distributed FFT libraries are evaluated: heFFTe-2.4.1~\cite{ayala2020heffte}, a high-performance FFT library, and FFT-MPI~\cite{plimpton2018fftmpi}, the default library in LAMMPS. The utofu-FFT leverages the BLAS library to perform the multiplication between twiddle factor matrices and real-space value, eliminating the need for FFTW. Each node launches four MPI ranks, with CPUs operating at 2.2 GHz in eco mode level 2.

Our optimized code is based on LAMMPS-22Dec22 and DeepMD-2.1.0. The baseline results in the paper correspond to the performance of the original code. The numbers of nodes in our test are 12, 96, 768, 1500, 4608, 8400, with the topology configuration to be $2\times3\times2$, $4\times6\times4$, $8\times12\times8$, $12\times15\times12$, $16\times18\times16$ and $20\times21\times20$ respectively.

\begin{figure*}[t]
  \centering
  \includegraphics[scale=1.0]{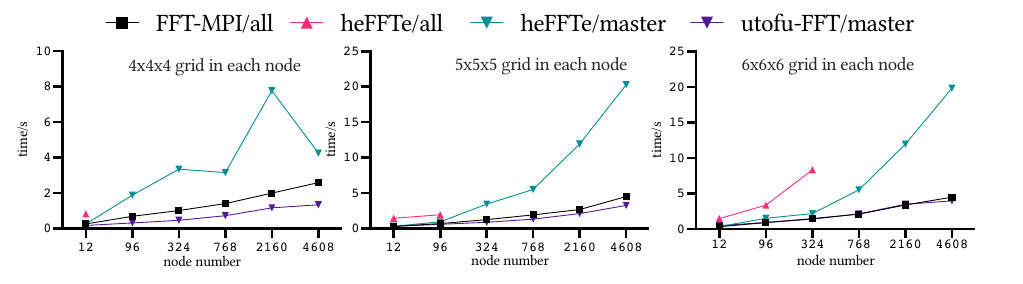}
  \caption{ 
  3D FFT test result. Three distributed FFT libraries are evaluated: FFT-MPI, heFFTe, and utofu-FFT. The benchmark measures the total execution time for 1,000 iterations of the \textit{brick2fft} and \textit{poisson} functions. Configurations with the suffix \textit{all} indicate that all MPI ranks participate in the FFT computation, while \textit{master} refers to configurations where only a single MPI rank per node performs the FFT computation.
  % 3D FFT实验测试结果。测试为brick2fft和poisson两个方程执行1000次迭代的时间。FFT-MPI和heFFTe以及utofu-FFT为分布式FFT计算库。尾缀all为所有进程参与FFT计算，master为一个节点中，只有一个进程参与FFT计算。
  }
  \label{fig:fft_result}
\end{figure*}

\subsection{Accuracy}

\begin{sloppypar}
We begin by conducting accuracy evaluations to investigate the influence of different numerical precision and 3D-FFT grid settings on simulation accuracy. The dataset used in this study is publicly available at https://doi.org/10.5281/zenodo.6024644~\cite{zhang2022deep}. The DP and DW models are trained for 1 million steps, and evaluation is performed on the final testing set. The evaluation is carried out on a 128-water molecular system, with the simulation box set to approximately 16~\AA~ in each direction and periodic boundary conditions. A total of 12 nodes are employed, where each node is assigned a subdomain of approximately size [8, 5.3, 8]\AA. It is noted that the box size varies slightly across cases in the test set, hence we use the term \textit{approximately}. Five different precision configurations are tested:
\end{sloppypar}

\begin{itemize}
    \item Double (baseline configuration): All computations are carried out in double precision with the FFT grid of [32, 32, 32]. This configuration is the baseline of our work.
    \item Mixed-FP32:  The precision of both FFT and neural network computations is reduced to single precision (float), while other components remain in double precision.
    \item Mixed-Int32: The reduction operations in the FFT computation are performed using 32-bit integer precision (int32). The FFT grid points are adjusted to [8, 12, 8], [10, 15, 10], or [12, 18, 12], ensuring that each node contains 4, 5, or 6 grid points along each direction.
\end{itemize}

Table~\ref{tab:error_energy} presents the error in single-step energy and force calculations, comparing DFT values with different precision configurations. The results show that our mixed-precision configurations maintain \textit{ab initio} accuracy. 
% Surprisingly, the error of the mixed-int2 configuration is exceptionally small, which we attribute to the dataset used.

We further conducted long-timescale simulations using \textit{double} and \textit{mixed-int2} precision, with the results presented in Fig.~\ref{fig:press}. It is clear that our optimized code closely aligns with the baseline results, demonstrating energy convergence during the MD simulations, which ensures both stability and accuracy.

\subsection{3D FFT optimization}

We evaluate three different grid configurations, where each node is assigned to $4\times4\times4$, $5\times5\times5$, and $6\times6\times6$ grids. Correspondingly, each MPI rank is assigned $2\times4\times2$, $2\times5\times2$ (or $3\times5\times3$), and $3\times6\times3$ grid points, respectively.

The tests are divided into two groups to analyze the impact of different FFT computation strategies on performance:

\begin{itemize}
    \item All MPI ranks participate in FFT computation: Every MPI rank performs 3D-FFT calculations using either the FFT-MPI or heFFTe library, corresponding to the \textbf{FFT-MPI/all} and \textbf{heFFTe/all} legends in Fig.~\ref{fig:fft_result}.
    \item Single-core for FFT computation per node: Only one core per node is designated for FFT computation, leveraging either the heFFTe or utofu-FFT library, corresponding to \textbf{heFFTe/master} and \textbf{utofu-FFT/master} legend in Fig.~\ref{fig:fft_result}.
\end{itemize}

We evaluate the total execution time of the \textit{brick2fft} and \textit{poisson\_ik} functions in LAMMPS, with each test running for 1,000 iterations. The \textit{brick2fft} function, which is only applicable to FFT-MPI, is responsible for transforming the grid distribution from brick to pencil. The \textit{poisson\_ik} function involves one forward 3D-FFT, three inverse 3D-FFTs, and a small portion of computation that takes negligible time. Therefore, the execution time of these two functions reflects the end-to-end efficiency of FFT computations.

The results are shown in Fig.~\ref{fig:fft_result}. In the first two grid configurations, the utofu-FFT achieves up to $2\times$ and $1.4\times$ speedup compared to FFT-MPI, primarily due to communication optimizations. Although the grid sizes are not powers of two, which typically leads to suboptimal performance in FFTW, the overhead is largely dominated by communication in these cases, making the impact of computational inefficiency negligible. In the last grid configuration, utofu-FFT is slightly outperforms FFT-MPI. This is because each node holds 216 grid points, requiring 36 reduction communications per node for a single FFT dimension when using utofu-FFT, introducing significant overhead. 

Additionally, heFFTe shows lower performance across all cases and lacks support for scenarios where each MPI rank has only a small number of grid points.

\subsection{Step-by-step optimization}

\begin{figure*}[htb]
  \centering
  \includegraphics[scale=0.7]{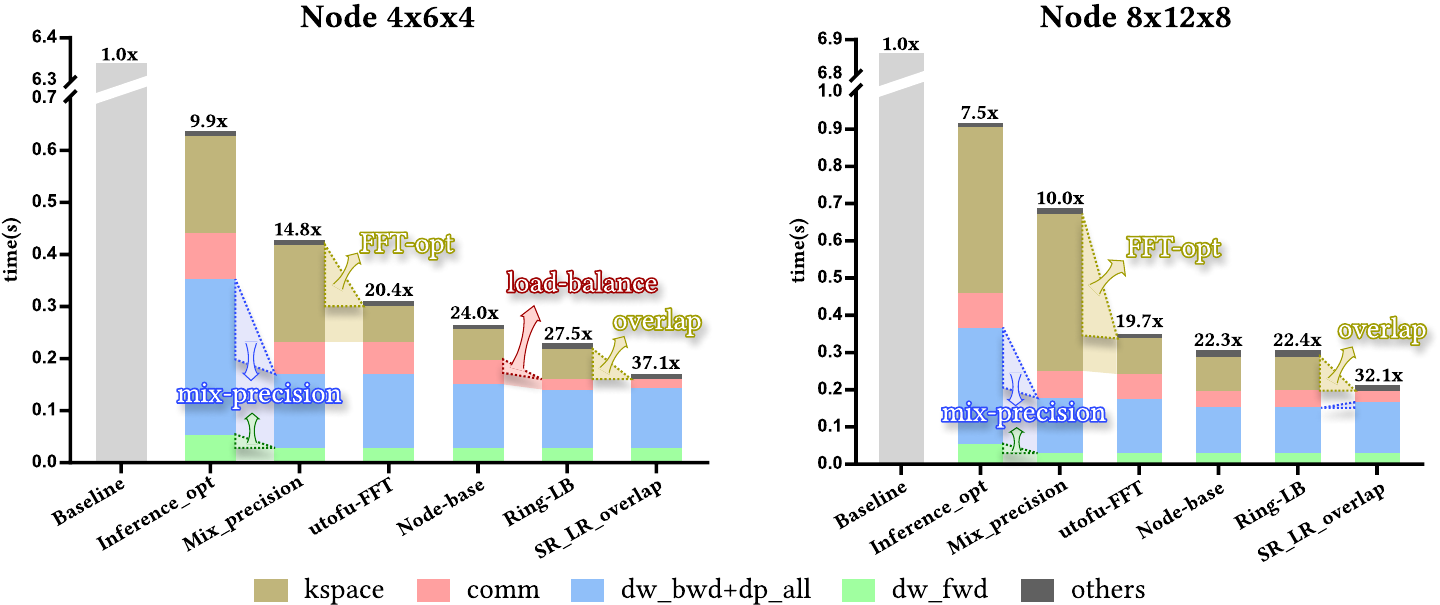}
  \caption{ 
  Step-by-step evaluation result for 100 time-steps. The \textit{kspace} label represents the time spent on the PPPM algorithm for long-range interactions, while \textit{comm} denotes the communication time within LAMMPS. The \textit{dw\_fwd} label corresponds to the forward phases of the DW model. \textit{dw\_bwd+dp\_all} labels indicate the total computation time for the DP model and DW model backpropagation phase. The \textit{others} bar includes all remaining operations, such as neighbor list construction, NVT ensemble updating (position and velocity integration), and system status output. The numbers above each bar indicate the speedup relative to the baseline configuration.
  }
  \label{fig:step-by-step}
\end{figure*}

We conduct detailed step-by-step performance evaluations on 96 and 768 nodes, maintaining an average of 47 atoms per node, equivalent to approximately one atom per core. The base simulation box size is 20.85\AA, containing 188 water molecules. For the 96-node and 768-node configurations, the base box is replicated twice and four times in each spatial direction, respectively. The results are shown in Fig.~\ref{fig:step-by-step}. The simulations run for 100 time-steps, plus a 10-step warm-up. Note that the baseline code is evaluated under 16 MPI ranks per node configuration, which yields the highest performance.

We first evaluate inference optimizations. By applying fine-grained kernel-level optimizations to the neural network and removing the TensorFlow framework, we achieve up to a $9.9\times$ and $7.5\times$ improvement in computational efficiency. Additionally, we observe that removing TensorFlow significantly reduces the application initialization overhead which is especially impactful in large-scale simulations where library loading can impose a substantial I/O burden. Furthermore, converting both FFT and neural network computations to single precision yields additional $1.5\times$ and $1.3\times$ speedup.

Integrating the utofu-FFT library leads to a $1.38\times$ and $2\times$ performance improvement on 96 nodes and 768 nodes, respectively. The larger gain at 768 nodes is attributed to the higher proportion of FFT time in the total runtime. Using the node-based decomposition method further improved performance by 18\% and 13\% by reducing communication overhead and improving intra-node load balance.

Next, after adding the ring-based load balancing scheme, the performance yields a 15\% gain on 96 nodes. In Fig.~\ref{fig:step-by-step} \textit{Ring-LB} bar, the improvement primarily stems from reduced communication time, while the reported force calculation time remains unchanged. This is because the force time reflects an average across all MPI ranks, which is not directly affected by load balance. However, due to data dependencies among MPI ranks in LAMMPS, load balance results in more uniform force computation times across MPI ranks, thereby reducing waiting time and communication delays. 

At 768 nodes, the improvement of load balancing is minimal. Further analysis revealed that, in some cases, the number of atoms an MPI rank needed to migrate for load balance exceeds its own atom count, making the ring-based load balancing scheme inapplicable, and then falling back to the intra-node load balance method. This limitation arises from the system generation strategy in LAMMPS, where large systems are created by replicating a base simulation box. If the original system is load imbalance, replication exacerbates it significantly. This issue can be mitigated in two ways: (1) by running long time-scale simulations to allow the system to equilibrate and naturally reduce the initial imbalance; or (2) by dividing nodes into smaller groups and applying ring-based balancing within each group to limit replication effects, though the latter only guarantees local load balance. 

Finally, we test the overlap of short-range and long-range force computations. At 96 nodes, the long-range computations are completely masked by short-range force, achieving a 35\% performance improvement. However, at 768 nodes, the overlap is incomplete, and the force computation time increased by 8\% compared to short-range force time alone. This is because the time for short-range forces remains nearly constant from 96 nodes to 768 nodes, while the long-range force computation nearly doubled, reaching the same level as short-range force time and limiting overlap effectiveness.

\begin{figure}[htb]
  \centering
  \includegraphics[scale=0.7]{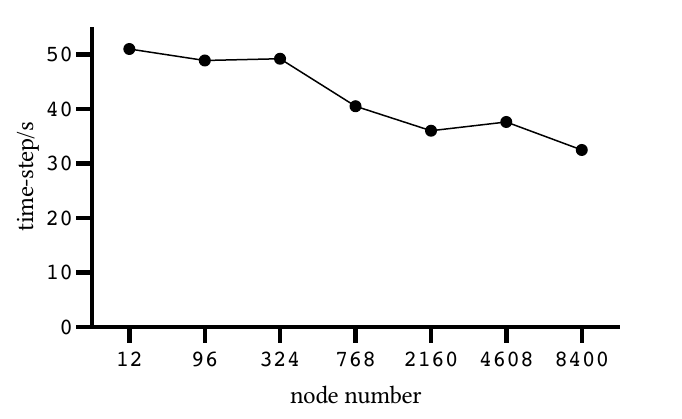}
  \caption{ Weak scaling evaluation, the number of nodes scales from 12 to 8,400, while maintaining a constant computational load of evenly 47 atoms per node.
  }
  \label{fig:strong_scaling}
\end{figure}

\subsection{Weak scaling result}

Fig~\ref{fig:strong_scaling} shows the weak scaling test result. The base box is the same as in the step-by-step evaluations. Simulations are run on 96, 324, 768, 2160, 4608 and 8400 nodes. The base simulation box is replicated by factors of (2,2,2), (3,3,3), (4,4,4), (6,5,6), (8,6,8), and (10, 7, 10) in the three spatial dimensions for each node configuration, respectively. The total atom number expands from 564 to 403,200. This ensures an average of 47 atoms per node throughout all configurations.

As the number of nodes increases, the proportion of long-range force computations gradually rises. Nevertheless, our optimized implementation maintains high computational efficiency. On 8,400 nodes, we achieve a simulation speed of 32.5 nanoseconds per day for a water system with 403,200 atoms.

\section{Conclusion}

In this work, we significantly improve the computational efficiency of the DPLR framework, substantially reducing the time-to-solution. We introduce uTofu-FFT, a hardware-offloaded distributed FFT scheme on Fugaku, achieving a $2\times$ speedup over the baseline FFT-MPI library. To further optimize performance, we propose a novel overlap strategy for long-range and short-range force computations that dedicates a single core per node to long-range interactions, effectively achieving near-complete overlap. Additionally, we develop a ring-based load balancing algorithm that enables global atom-level evenly redistribution with minimal communication overhead. Our optimizations yield a $37\times$ speedup over the baseline, achieving 51 ns/day for a 564-atom water system on 12 nodes and  32 ns/day for a 400K-atom water system on 8,400 nodes.

\begin{acks}
 This work is supported by the following funding: National Key Research and Development Program of China (2021YFB0300600), National Science Foundation of China (92270206, T2125013, 62372435, 62032023, 61972377,  61972380, T2293702), CAS Project for Young Scientists in Basic Research (YSBR-005) and the Strategic Priority Research Program of Chinese Academy of Sciences, Grant No. XDB0500102, China National Postdoctoral Program for Innovative Talents (BX20240383), and Huawei Technologies Co., Ltd.. This work used computational resources of the supercomputer Fugaku provided by RIKEN through the HPCI System Research Project (Project ID: hp240303). The numerical calculations in this study were partially carried out on the ORISE Supercomputer. We thank Prof. Linfeng Zhang  (Peking University) for helpful discussions.
\end{acks}

% \bibliographystyle{ACM-Reference-Format}
% \bibliography{section/reference}


\end{document}